\documentclass[12pt]{article}
\usepackage{amssymb,amsmath,epsfig}
\allowdisplaybreaks

\begin{document}

\title{\bf Stability of Charged Thin-Shell Gravastars with Quintessence}
\author{M. Sharif \thanks {msharif.math@pu.edu.pk} and Faisal Javed
\thanks{faisalrandawa@hotmail.com}\\
Department of Mathematics, University of the Punjab,\\
Quaid-e-Azam Campus, Lahore-54590, Pakistan.}

\date{}
\maketitle

\begin{abstract}
This paper develops a new solution of gravitational vacuum star in
the background of charged Kiselev black holes as an exterior
manifold. We explore physical features and stability of thin-shell
gravastars with radial perturbation. The matter thin layer located
at thin-shell greatly affects stable configuration of the developed
structure. We assume three different choices of matter distribution
such as barotropic, generalized Chaplygin gas and generalized
phantomlike equation of state. The last two models depend on the
shell radius, also known as variable equation of state. For
barotropic model, the structure of thin-shell gravastar is mostly
unstable while it shows stable configuration for such type of matter
distribution with extraordinary quintessence parameter. The
resulting gravastar structure indicates stable behavior for
generalized Chaplygin gas but unstable for generalized phantomlike
model. It is also found that proper length, entropy and energy
within the shell show linear relation with thickness of the shell.
\end{abstract}

\section{Introduction}

The study of final outcomes of gravitational collapse is an
interesting topic that explores the formation of various compact
objects such as white dwarfs, neutron stars, naked singularities and
black holes (BHs). The collapse end-state is a widely accepted
research field from many perspectives, both theoretical and
observational. The classical general relativity faces some major
scientific issues precisely related to the paradoxical
characteristics of BHs and naked singularities. An astronomical
substance hypothesized as a substitute for the BH is a gravastar
(gravitational vacuum star) based on the idea of Mazur's and
Motola's theory \cite{1,2}. The basic idea is to prevent the
formation of event horizons and singularities by stopping the
collapse of matter at or near the position of event horizon. A
gravastar appears similar to a black hole but does not contain event
horizon and singularity.

Gravastars are of purely theoretical interest and can be described
in three different regions with the specific equation of state
(EoS). The first region is referred to as an interior ($0\leq
r<r_1$), second is the intermediate ($r_1< r<r_2$) and third is
denoted as an exterior region ($r_2<r$). In the first region, the
isotropic pressure ($p=-\sigma$, where $\sigma$ represents the
energy density) produces a repulsive force on the intermediate
thin-shell. It is assumed that the intermediate thin-shell is
protected by ultrarelativistic plasma and fluid pressure
($p=\sigma$). The exterior region has zero pressure ($p=0=\sigma$)
and can be supported by the vacuum solution of the field equations.
It contains a stable thermodynamic solution and maximum entropy for
small fluctuations. Visser's cut and paste method provides a general
formalism for the construction of thin-shell from the joining of two
different spacetimes at hypersurface \cite{2a}. Mazur and Mottola
\cite{2} considered this approach to construct thin-shell gravastar
from the matching of exterior Schwarzschild BH with interior de
Sitter (DS) spacetime. This approach is very useful to avoid the
presence of event horizon as well as central singularity in the
geometry of gravastars.

The matter surface at thin-shell creates a sufficient amount of
pressure to overcome the force of gravity effects that help to
maintain its stable configuration. For the description of
Mazur-Mottola scenario, Visser and Wiltshire \cite{3} introduced the
simplest model from the matching of exterior and interior geometries
through the cut and paste approach. They also analyzed stable
configuration of the developed structure for suitable choice of EoS
for the transition layers. Carter \cite{4} extended this concept by
the joining of interior DS spacetime and exterior
Reissner-Nordstr\"om (RN) BH. They examined the effects of EoS on
the modeling of thin-shell gravastars. Horvat et al. \cite{5}
presented theoretical model of gravastars with electromagnetic field
and studied the role of charge on the stable configuration of
gravastars. Rahaman et al. \cite{6} studied physical features like
proper length, entropy and energy contents of charged and charged
free thin-shell gravastars in the background of (2+1)-dimensional
spacetime. They claimed that the presented solutions are
non-singular and physically viable as an alternative to BH. Banerjee
et al. \cite{7} investigated the braneworld thin-shell gravastars
developed by using braneworld BH as an exterior manifold through cut
and paste technique.

Rocha and his collaborators \cite{8} discussed stable configuration
of thin-shell gravastars with perfect fluid distribution in Vaidya
exterior spacetime. They proposed a dynamical model of prototype
gravastars filled with phantom energy. It is found that the
developed structure can be a BH, stable, unstable or ``bounded
excursion" gravastar for various matter distributions at thin-shell.
Horvat et al. \cite{9} studied the geometry of gravastars with
continuous pressure by using the conventional Chandrasekhar approach
and derived EoS for the static case. Lobo and Garattini \cite{10}
investigated the stability of noncommutative thin-shell gravastar
and found that stable regions must exist near the expected position
of the event horizon. \"{O}vg\"{u}n et al. \cite{11} developed
thin-shell gravastar model from the matching of exterior charged
noncommutative BH with interior DS manifold. They found that the
developed structure follows the null energy condition and shows
stable behavior for some suitable values of physical parameter near
the expected event horizon. Recently, we have developed regular
thin-shell gravastars in the background of Bardeen/Bardeen DS BHs as
exterior manifolds through cut and paste method \cite{12}. The
stable configuration of the developed structure is explored through
radial perturbation. It is found that stable regions decrease for
large values of charge and increase for higher values of the
cosmological constant.

The theoretical modeling of gravastar could be helpful for the
better understanding of dark energy role in the accelerated
expanding behavior of the universe. Ghosh et al. \cite{13} examined
physical characteristics of gravastar model with Kuchowicz metric
potential. They claimed that this model overcomes the singularity
problems that occurred for the geometry of BH in general relativity.
Shamir and Ahmad \cite{14} investigated physical features of
gravastar model in the background of $f(G,T)$ gravity. Yousaf et al.
\cite{16} explored stable configuration of charged gravastar filled
with isotropic fluid in $f(R,T)$ gravity. They found linear relation
among the physical features and thickness of the shell. Sharif and
Waseem \cite{17} discussed charged gravastars with conformal motion
in $f(R,T)$ gravity. There is a large body of literature
\cite{18}-\cite{31a} that explore the stable as well as dynamical
configuration of thin-shell wormholes constructed from the matching
various BHs with different EoS.

This paper presents the formalism of charged Kiselev thin-shell
gravastars to explore stable configuration with different EoS. The
paper has the following format. Section \textbf{2} explains the
formalism of thin-shell gravastars in the background of charged
Kiselev BH. In section \textbf{3}, we study the effects of
barotropic and variable EoS on the stable configuration of the
developed structure through radial perturbation. Finally, we
summarize our results in the last section.

\section{Gravastars Formalism}

This section explores the geometrical construction of thin-shell
gravstars from the joining of lower ($\Upsilon^-$) and upper
($\Upsilon^+$) manifolds through cut and paste technique. For this
purpose, we consider DS spacetime as a lower manifold and charged BH
surrounded by the quintessence matter as an upper manifold. The
motivation behind the consideration of this model can be explained
as follows. The matter with negative pressure can be characterized
for the current evolutionary phase of the universe with cosmological
constant and quintessence \cite{31b}. The mathematical
representation of quintessence matter distribution that linear
relates energy density ($\sigma_q$) and pressure ($p_q$) is
$p_q=w\sigma_q$, where $\omega$ denotes the quintessence parameter.
This parameter explains that the universe is in the phase of
accelerated expansion if $-1<\omega<-1/3$, decelerates if
$\omega>-1/3$ and shows inertial behavior (constant expansion rate)
if $\omega=-1/3$. This means that observers must have future
horizons in all accelerated models \cite{31c}. In an accelerated
expanding universe, two objects separated with a relative fixed
distance $r$ must achieve relative speed to the speed of light after
some time and will no longer communicate. For the case of
decelerated expansion, the breakdown of such a communication does
not happen whereas it becomes less relativistic with time. However,
the speed of relative moving observers must be constant for the case
of $\omega=-1/3$. They can communicate but cannot maintain this
forever as they recede away from each other.

Kiselev \cite{32} introduced uncharged and charged BH surrounded by
the quintessence matter distribution as a static spherically
symmetric solution of the field equations. The line element of
charged Kiselev BH is given as
\begin{equation}\label{1}
ds^2_+=-\Psi(r_+)dt^{2}+\Psi(r_+)^{-1}dr^{2}_++r^2_+d\theta^2_++
r^2_+\sin^2\theta_+ d\phi^2_+,
\end{equation}
where
\begin{equation}\nonumber
\Psi(r_+)=1-\frac{2m}{r_+}-\frac{\alpha}
{r^{3\omega+1}_+}+\frac{Q^2}{r^2_+},
\end{equation}
$m$ is the mass of BH, $Q$ denotes the charge of BH, $\alpha$ stands
for the Kiselev parameter and $\omega$ is the quintessence parameter
with $-1<\omega<-1/3$. The boundary values of EoS parameter recover
the case of cosmological constant (extraordinary quintessence) for
$\omega=-1$ and $\omega=0$ is referred to as the dust fluid. If
$Q=0$, then it reduces to Kiselev BH and RN BH is recovered when
Kiselev parameter vanishes. The charged Kiselev BH reduces to
Schwarzschild BH in the absence of both charge and Kiselev
parameter. The corresponding metric function of Kiselev BH has the
following form
\begin{equation}\nonumber
\Psi(r_+)=1-\frac{2m}{r_+}-\frac{\alpha} {r^{3\omega+1}_+}.
\end{equation}

Extreme BHs are expected to have both stable and unstable
properties, this makes their analysis very interesting and
challenging. We consider $\omega=-2/3\in(-1,-1/3)$ to observe the
event horizon of Kiselev BH. The corresponding event horizons are
given as
\begin{equation}\nonumber
r_h=\frac{1\pm\sqrt{1-8 \alpha  m}}{2 \alpha }.
\end{equation}
It is found that
\begin{itemize}
\item for $\alpha=1/8m$, it denotes extreme Kiselev BH,
\item for $\alpha<1/8m$, it represents the non-extreme Kiselev BH,
\item for $\alpha>1/8m$, it shows naked singularity.
\end{itemize}
Since the charged Kiselev BH metric function is much complicated
than RN and Kiselev BH, so its event horizon for $\omega=-2/3$ has
much complicated expression. Thus we only discuss values of the
parameter for which it shows different geometrical structure. It
follows that
\begin{itemize}
\item for $Q^2=\frac{2}{27\alpha^2}\left(-2+18m\alpha-2(1-6m\alpha)^{3/2}\right)$,
it denotes extreme charged Kiselev BH,
\item for $Q^2>\frac{2}{27\alpha^2}\left(-2+18m\alpha-2(1-6m\alpha)^{3/2}\right)$,
it represents the non-extreme charged Kiselev BH,
\item for $Q^2<\frac{2}{27\alpha^2}\left(-2+18m\alpha-2(1-6m\alpha)^{3/2}\right)$,
it shows naked singularity.
\end{itemize}

The line element of DS geometry is given as
\begin{equation}\label{2}
ds^2_{-}=-\Phi(r_{-})dt_{-}^2+\Phi^{-1}(r_{-})dr_{-}^2
+r_{-}^2d\theta_{-}^2+r_{-}^2\sin^2\theta_{-} d\phi_{-}^2,
\end{equation}
where $\Phi(r_{-})=1-r^2_-/\beta^2$ and $\beta$ is a nonzero
positive constant. We use cut and paste method to obtain the
geometry of thin-shell gravastars from the matching of two distinct
spacetimes $\Upsilon^-$ and $\Upsilon^+$. These manifolds have the
metric functions defined by $g^{\pm}_{\mu\nu}(x_{\pm}^\mu)$ with
independent coordinates $x_{\pm}^\mu$ and bounded by the
hypersurfaces $\partial\Upsilon^{\pm}$ with induced metrics
$h_{ij}^{\pm}$, respectively. According to the Darmoise junction
conditions, the induced metrics are isometric and follow the
relation $h_{ij}^{+}(\xi^i)=h_{ij}(\xi^i) =h_{ij}^{-}(\xi^i)$, where
$\xi^i$ represents the coordinates of $\partial\Upsilon^{\pm}$.
These geometries are glued at the hypersurface to obtain the single
manifold $\Upsilon=\Upsilon^+\cup\Upsilon^-$ with boundary
$\partial\Upsilon=\partial\Upsilon^+=\partial\Upsilon^-$.
Mathematically, these spacetimes can be described as
\begin{equation}\nonumber
\Upsilon^\pm=\{x_{\pm}^\mu|t_{\pm}\geq T_{\pm}(\tau)
\quad\text{and}\quad r\geq b(\tau)\},
\end{equation}
where $\tau$ and $b(\tau)$ denote the proper time and radius of
thin-shell. The corresponding hypersurface that linked these
geometries can be parameterized as
\begin{equation}\nonumber
\partial\Upsilon=\{\xi^i|t_{\pm}\geq T_{\pm}(\tau)
\quad\text{and}\quad r=b(\tau)\}.
\end{equation}

The induced 3D metric at hypersurface $(h_{ij})$ can be expressed as
\begin{equation}\nonumber
ds^2_{\partial\Upsilon}=h_{ij}d\xi^id\xi^j=-d\tau^2+b(\tau)^2d
\theta^2+b(\tau)^2 \sin^2\theta d\phi^2,
\end{equation}
where $\xi^{i}=(\tau,\theta,\phi)$. The normal vector components of
$g_{\mu\nu}$ on the $\partial\Upsilon$ are defined as
\begin{equation}\nonumber
n^{\mu}=\frac{f(r,b(\tau))_{,\mu}}{|f(r,b(\tau))_{,\nu}
f(r,b(\tau))^{,\nu}|^{1/2}},
\end{equation}
where $f(r,b(\tau))=r-b(\tau)=0$ represents the function of
$\partial\Upsilon$ and $b(\tau)=b$ denotes the shell's radius. The
components of normal vectors corresponding to upper and lower
spacetimes are
\begin{eqnarray}\label{3}
n_{+}^{\mu}&=&\left(\frac{\dot{b}}{1-\frac{2m}{b}-\frac{\alpha}
{b^{3\omega+1}}+\frac{Q^2}{b^2}},\sqrt{1-\frac{2m}{b}-\frac{\alpha}
{b^{3\omega+1}}+\frac{Q^2}{b^2} +\dot{b}^2},0,0\right),\\\label{4}
n_{-}^{\mu}&=&\left(\frac{\dot{b}}{1-\frac{b^2}{\beta^2}},\sqrt{1-\frac{b^2}{\beta^2}
+\dot{b}^2},0,0\right),
\end{eqnarray}
respectively. Here, dot represents derivative with respect to
$\tau$. The normal vector satisfies the condition $n^\mu n_\mu=1$
for spherical symmetric manifolds. The discontinuity in the second
fundamental form (extrinsic curvature) exist due to the presence of
matter surface at $\partial\Upsilon$. The extrinsic curvature
components for both geometries are
\begin{eqnarray}\label{5}
K_{\tau}^{\tau+}&=&\frac{\alpha(3 \omega +1) b^{1-3 \omega }+2 b m-2
Q^2+2\ddot{b}b^3}{b^3\sqrt{1-\frac{2m}{b}-\frac{\alpha}
{b^{3\omega+1}}+\frac{Q^2}{b^2}+\dot{b}^2}},\\\label{6}
K_{\theta}^{\theta+}&=&
\frac{1}{b}\sqrt{1-\frac{2m}{b}-\frac{\alpha}
{b^{3\omega+1}}+\frac{Q^2}{b^2}+\dot{b}^2},\\\label{7}
K_{\tau}^{\tau-}&=&\frac{-\frac{2 b}{\beta ^2}+2\ddot{b}}
{\sqrt{1-\frac{b^2}{\beta^2}+\dot{b}^2}},\\\label{8}
K_{\theta}^{\theta-}&=& \frac{1}{b}\sqrt{
1-\frac{b^2}{\beta^2}+\dot{b}^2},\\\label{9}
K^{\phi\pm}_{\phi}&=&\sin^2\theta K_{\theta}^{\theta\pm},
\end{eqnarray}

The matter surface at thin-shell produces discontinuity in the
extrinsic curvatures of both spacetimes. If
$K_{ij}^{+}-K_{ij}^{-}\neq0$, then it represents the presence of
matter thin layer on $\partial\Upsilon$. The components of
energy-momentum tensor ($S^{i}_{j}$) of such a matter surface are
determined by the Lanczos equations. Mathematically, it can be
expressed as
\begin{equation}\label{10}
S^{i}_{j}=-\frac{1}{8\pi}\{[K^{i}_{j}]-\delta^{i}_{j}K\},
\end{equation}
where $[K^{i}_{j}]=K^{+i}_{j}-K^{-i}_{j}$ and
$K=tr[K_{ij}]=[K^{i}_{j}]$. The above equation in terms of perfect
fluid distribution becomes
\begin{equation}\label{11}
S^{i}_{j}=v^i v_j\left(p+\sigma\right)+p\delta^{i}_{j},
\end{equation}
here $v_i$ denotes thin-shell velocity components. By considering
Eqs.(\ref{5})-(\ref{11}), we obtain $\sigma$ and $p$ in the
following form
\begin{eqnarray}\label{12}
\sigma&=&-\frac{1}{4\pi b}\left\{\sqrt{1-\frac{2m}{b}-\frac{\alpha}
{b^{3\omega+1}}+\frac{Q^2}{b^2}+\dot{b}^2}-\sqrt{1-\frac{b^2}{\beta^2}+\dot{b}^2}\right\},
\\\label{13}
p&=&\frac{2\dot{b}^2+2b\ddot{b}+\frac{\alpha  (3 \omega -1) b^{-3
\omega }-2 m}{b}+2}{8\pi b\sqrt{1-\frac{2m}{b}-\frac{\alpha}
{b^{3\omega+1}}+\frac{Q^2}{b^2}+\dot{b}^2}}-\frac{2\dot{b}^2+2b\ddot{b}
+2-\frac{4 b^2}{\beta ^2}}{8\pi
b\sqrt{1-\frac{b^2}{\beta^2}+\dot{b}^2}}.
\end{eqnarray}
Here, we assume that $\dot{b}_{0}=\ddot{b}_{0}=0$, where $b_0$ is
the position of equilibrium shell's radius. This shows that shell's
motion along the radial direction vanishes at $b=b_0$. The
respective expressions for surface stresses at $b=b_0$ yield
\begin{eqnarray}\label{14}
\sigma(b_0)=\sigma_0&=&-\frac{1}{4\pi
b_0}\left\{\sqrt{1-\frac{2m}{b_0}-\frac{\alpha}
{b_0^{3\omega+1}}+\frac{Q^2}{b_0^2}}-\sqrt{1-\frac{b^2_0}{\beta^2}}\right\},
\\\label{15}
p(b_0)=p_0&=&\frac{1}{8\pi b_0}\left\{\frac{\alpha  (3 \omega -1)
b_0^{-3 \omega }-2 m+2b_0} {b_0\sqrt{1-\frac{2m}{b_0}-\frac{\alpha}
{b^{3\omega+1}_0}+\frac{Q^2}{b^2_0}}}-\frac{2\beta ^2-4 b_0^2}{\beta
^2\sqrt{1-\frac{b^2_0}{\beta^2}}}\right\}.
\end{eqnarray}
The continuity of perfect fluid gives the relationship between the
surface stresses of thin-shell gravastars as
\begin{eqnarray}\label{16}
4\pi\frac{d}{d\tau}( b^2 \sigma)+4\pi p\frac{d b^2}{d\tau}=0,
\end{eqnarray}
which can be expressed as
\begin{eqnarray}\label{17}
\frac{d\sigma}{db}=-\frac{2}{b}(\sigma+p).
\end{eqnarray}
The second order derivative of $\sigma$ with respect to $b$ yields
\begin{equation}\label{18}
\frac{d^2\sigma}{db^2}=\frac{2(p+\sigma)}{b^2}\left(3+2\varsigma^2\right),
\end{equation}
where $\varsigma^2=dp/d\sigma$ denotes the EoS parameter. Equations
(\ref{16})-(\ref{18}) are very useful to explore the dynamics and
stable configurations of constructed geometry with different types
of matter distribution.

\begin{figure}\centering
\epsfig{file=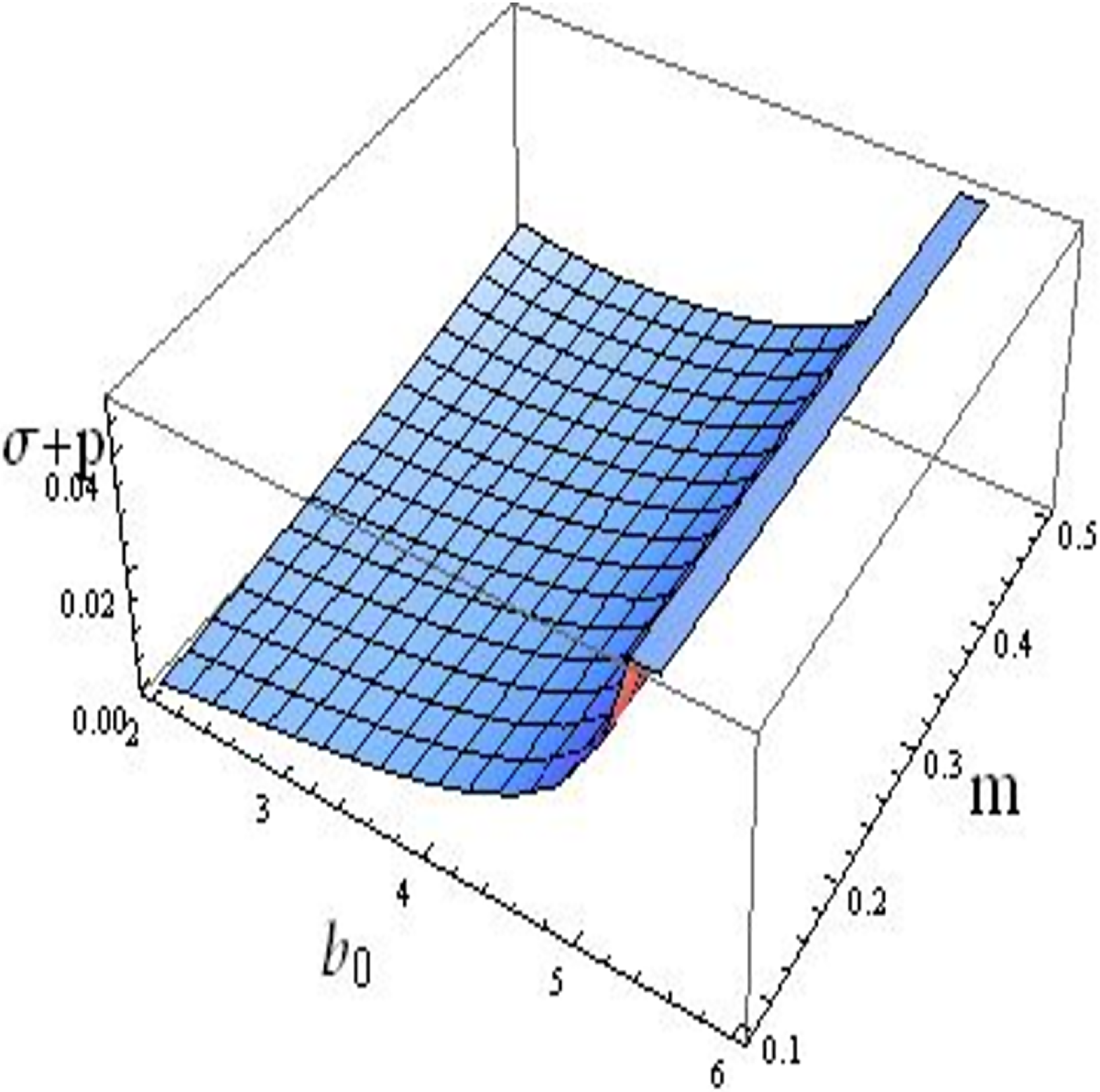,width=0.5\linewidth}\epsfig{file=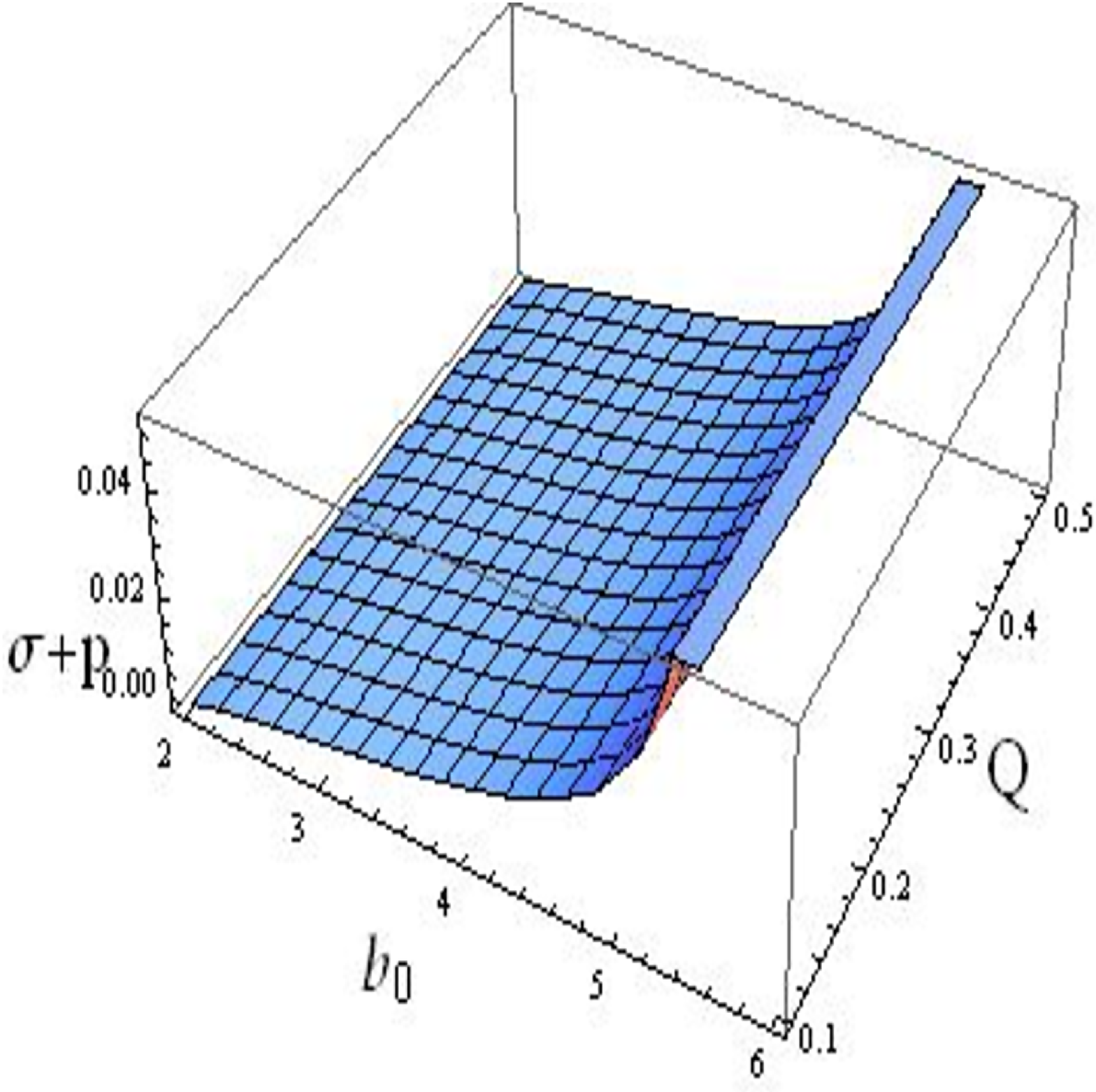,width=0.5\linewidth}
\caption{Plots of the null energy condition for charged Kiselev
thin-shell gravatars. We examine graphical behavior of $\sigma+p$ at
$b=b_0$ verses $(b_0,Q)$ (left plot) and $(b_0,m)$ (right plot) for
$\omega=-2/3$, $\alpha=0.2$ and $\beta=0.5$.}\label{f1}
\end{figure}

For the physical viability of a geometrical structure, some
constraints must be imposed known as energy conditions. The
well-known energy conditions are null: $\sigma_0+p_0>0$; weak:
$\sigma_0>0$, $\sigma_0+p_0>0$; strong: $\sigma_0+3p_0>0$,
$\sigma_0+p_0>0$; dominant: $\sigma_0>0$, $\sigma_0\pm p_0>0$. If
these energy conditions are verified then the developed model is
physically viable. Here, we are interested to check the null energy
condition that ensure the presence of normal or exotic matter at
thin-shell. It is interesting to mention here that the violation of
the null energy condition leads to the violation of remaining
conditions. We see that thin-shell gravastars follow the null energy
condition for different values of charge and mass of BH as shown in
Figure \ref{f1}. These values of physical parameters have
frequently been used in literature that examine the stable as well
as dynamical behavior of thin-shell constructed from different
singular and non-singular BHs \cite{20}-\cite{31a}. Thus we use them
to determine the effects of charge and mass on the energy
conditions, physical features as well as stability of thin-shell
gravastars (see Appendix \textbf{A}).

\section{Stability Analysis}

This section studies stability of thin-shell gravastars using linear
perturbation in the radial direction at $b=b_0$ with different
variable EoS. The stable and unstable configurations of thin-shell
gravastars can be analyzed through the behavior of effective
potential of thin-shell. The equation of motion of thin-shell that
explains the stable as well as dynamical characteristics of
respective geometry is obtained directly from Eq.(\ref{12}) as
\begin{equation}\label{26}
\dot{b}^2+\Omega(b)=0,
\end{equation}
here $\Omega(b)$ denotes the potential function of thin-shell
gravastar as
\begin{equation}\label{27}
\Omega(b)=-\frac {\xi(b)^2}{64 \pi ^2 b^2 \sigma
^2}+\frac{1}{2}\zeta(b)-4 \pi ^2 b^2\sigma ^2,
\end{equation}
where
\begin{equation}\nonumber
\xi(b)=\frac{\alpha  b^{1-3 \omega }-\frac{b^4}{\beta ^2}+2 b
m-Q^2}{b^2}, \quad \zeta(b)=2+\frac{-\alpha  b^{1-3 \omega
}-\frac{b^4}{\beta ^2}-2 b m+Q^2}{b^2}.
\end{equation}

The stable behavior of thin-shell gravastars is studied by using
second derivative of the effective potential at $b=b_0$. The basic
conditions for the stable behavior can be written as
$\Omega(b_0)=0=\Omega'(b_0)$ and $\Omega''(b_0)>0$. If
$\Omega''(b_{0})<0$, then it shows unstable behavior and it is
unpredictable if $\Omega''(b_{0})=0$ \cite{27}. To check the
stability through radial perturbation, we linearize the potential
function using Taylor series expansion around equilibrium radius
$b_0$ as follows
\begin{equation}\nonumber
\Omega(b)=\Omega(b_{0})+\Omega'(b_{0})(b-b_{0})+\frac{1}{2}
\Omega''(b_{0})(b-b_{0})^2+O[(b-b_{0})^3].
\end{equation}
We examine that $\Omega(b_0)=0=\Omega'(b_0)$, hence it reduces to
\begin{equation}\label{28}
\Omega(b)=\frac{1}{2}(b-b_{0})^2\Omega''(b_{0}).
\end{equation}
The corresponding second derivative of $\Omega(b)$ at $b=b_0$
becomes
\begin{eqnarray}\nonumber \Omega''(b_0)&=&\frac{2
M(b_0) M'(b_0)}{b^3_0}-\frac{b_0^2 \xi(b_0)\xi''(b_0)}{2M(b_0)^2}
+\frac{2 b_0^2 \xi(b_0) M'(b_0) \xi'(b_0)} {M(b_0)^3}
\\\nonumber&-&\frac{M'(b_0)^2} {2b_0^2}-\frac{b_0^2
\xi'(b_0)^2}{2M(b_0)^2}-\frac{3 b_0^2 \xi(b_0)^2
M'(b_0)^2}{2M(b_0)^4}-\frac{2 b_0 \xi(b_0) \xi'(b_0)}
{M(b_0)^2}\\\nonumber&+&\frac{\zeta''(b_0)}{2}-\frac{3
M(b_0)^2}{2b_0^4}+\frac{ b_0^2 \xi(b_0)^2 M''(b_0)}
{2M(b_0)^3}-\frac{M(b_0) M''(b_0)}{2b_0^2} \\\label{29}&
-&\frac{\xi(b_0)^2}{2M(b_0)^2}+\frac{2 b_0 \xi(b_0)^2
M'(b_0)}{M(b_0)^3},
\end{eqnarray}
where $M(b_0)=4\pi b_0^2 \sigma_0$ denotes the total mass
distribution at equilibrium shell's radius. The corresponding first
and second derivatives of the total mass with respect to $b$ at
$b=b_0$ become
\begin{eqnarray}\nonumber M'(b_0)=-8\pi b_0 p_0,\quad
M''(b_0)=-8\pi p_0+16\pi \varsigma_0^2 (\sigma_0+p_0),
\end{eqnarray}
and $\varsigma_0^2=dp/d\sigma|_{b=b_0}$.

\begin{figure}\centering
\epsfig{file=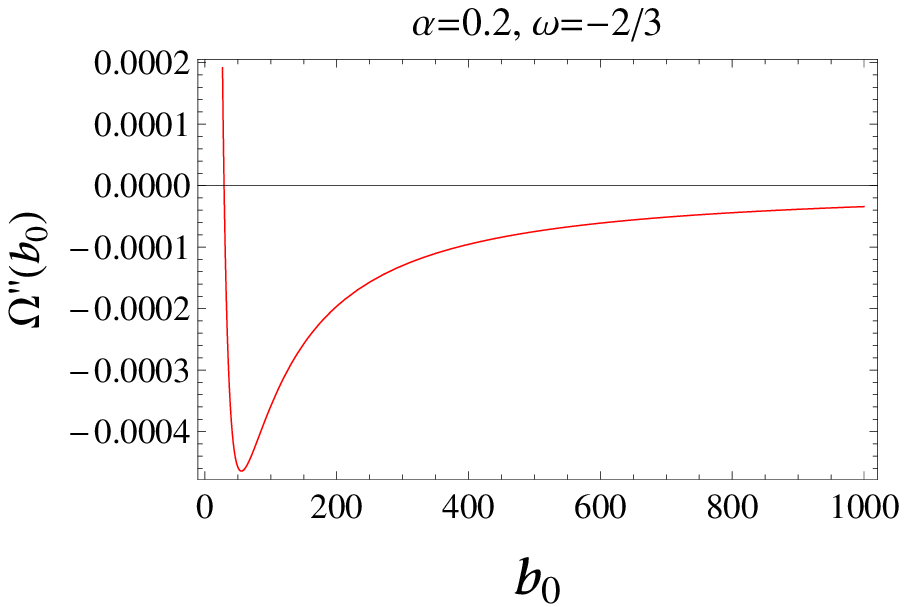,width=.5\linewidth}\epsfig{file=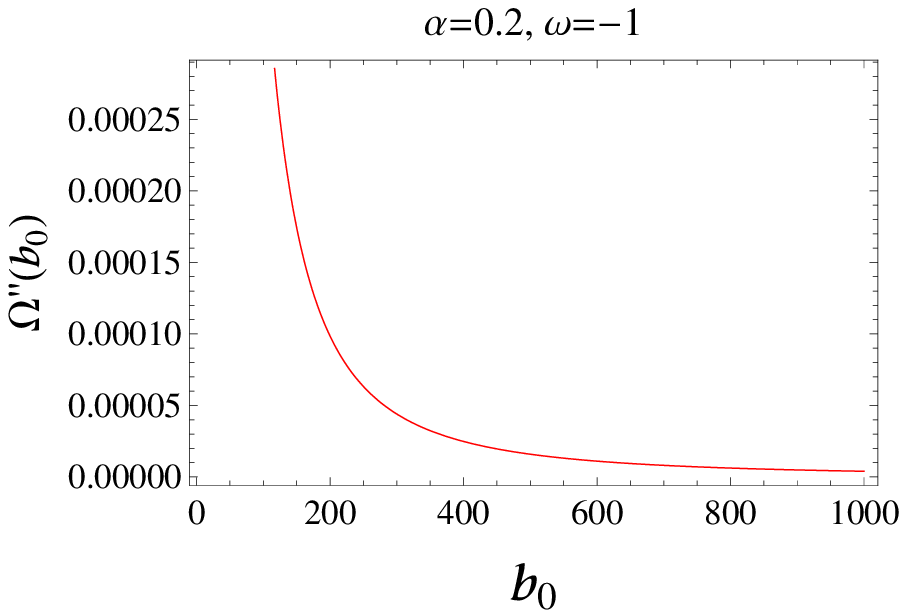,width=.5\linewidth}
\caption{Stability of thin-shell gravastars with barotropic EoS for
$\beta=0.5=m=Q$ with different values of $\omega$. The left plot
shows unstable behavior and right plot expresses the stable structure.}\label{f2}
\end{figure}

Firstly, we begin with barotropic EoS to discuss the stability of
the developed geometry. It gives linear relation between the surface
stresses of thin-shell as $p=\gamma \sigma$ with real constant
$\gamma$. Consequently, the solution of conservation equation
(\ref{17}) for barotropic EoS is given as
\begin{equation}\label{30}
\sigma=\left(b_{0}b^{-1}\right)^{2(1+\gamma)}\sigma_{0}.
\end{equation}
The corresponding potential function becomes
\begin{equation}\label{31}
\Omega(b)=\frac{\zeta(b)}{2}-\frac{\xi(b)^2
\left(\frac{b_0}{b}\right)^{-4 (\gamma+1)}}{64 \pi ^2 b^2
\sigma_0^2}-4 \pi ^2 b^2 \sigma_0^2 \left(\frac{b_0}{b}\right)^{4
(\gamma+1)},
\end{equation}
which turns out to be zero at throat radius $b=b_0$. The
corresponding first derivative of $\Omega(b)$ yields
\begin{equation}\label{32}
\Omega'(b_0)=\frac{\zeta'(b_0)}{2}-\frac{\xi(b_0) \left(2 \gamma
\xi(b_0)+b_0 \xi'(b_0)+\xi(b_0)\right)}{32 \pi ^2 \sigma_0^2
b_0^3}+8 \pi ^2 \sigma_0^2 (2 \gamma+1) b_0,
\end{equation}
which vanishes only if
\begin{equation}\label{33}
\gamma=\frac{-256 \pi ^4 \sigma_0^4 b_0^4-16 \pi ^2 \sigma_0^2 b_0^3
\zeta'(b_0)+b_0 \xi(b_0)\xi'(b_0)+\xi(b_0)^2}{2 \left(256 \pi ^4
\sigma_0^4 b_0^4-\xi(b_0)^2\right)}.
\end{equation}
The second derivative of $\Omega(b)$ at $b=b_0$ yields
\begin{eqnarray}\nonumber
\Omega''(b_0)&=&\frac{\zeta''(b_0)}{2}-\frac{1}{32 \pi ^2 \sigma_0^2
b_0^4}\left\{b_0 \xi(b_0) \left((8 \gamma+4)
\xi'(b_0)+b_0\xi''(b_0)\right)\right.\\\nonumber&+&\left.\left(8
\gamma^2+6 \gamma+1\right) \xi(b_0)^2+b_0^2
\xi'(b_0)^2\right\}\\\label{34}&-&8 \pi ^2 \sigma_0^2 (2 \gamma+1)
(4 \gamma+3),
\end{eqnarray}
This equation explains stable and unstable configurations of
thin-shell gravastars for barotropic EoS. Due to complexity of this
expression, we use numerical approach to observe the effects of
physical parameters on the stability of developed structure. We
study the graphical behavior of $\Omega''(b_{0})$ by using
Eqs.(\ref{33}) and (\ref{14}). It is found that stable structure of
thin-shell is greatly affected by the presence of quintessence EoS
parameter. We examine that thin-shell expresses unstable behavior
for every values of $Q$, $m$, $\alpha$ and $\beta$ with
$\omega=-2/3$ as shown in the left plot of Figure \ref{f2}. We
obtain unstable configuration for every choice of $\omega$ except
for extraordinary quintessence parameter $\omega=-1$ (right plot of
Figure \ref{f2}). Hence, the barotropic type fluid distribution at
thin-shell shows stable behavior only for $\omega=-1$ otherwise
gives unstable solutions.

Current observational data seem to point towards an accelerated
expansion of the universe \cite{32a}. If general relativity is
assumed to be correct theory of gravity describing the large-scale
behavior of the universe, then its energy density and pressure
should violate the strong energy condition. Several models for the
matter leading to such a situation have been proposed \cite{32b}.
One of them is the Chaplygin gas \cite{32c}, a perfect fluid
satisfying the EoS $p \sigma= \eta$, where $\eta<0$. A remarkable
property of the Chaplygin gas is that the squared sound velocity
$v_s^2=\eta/\sigma^2$ is always positive even in the case of exotic
matter. Varela \cite{18} considered the EoS of the type
$p=p(\sigma,b)$ to discuss the stability of thin-shell wormhole
developed from two equivalent copies of the Schwarzschild BH. Such
type of EoS is known as variable EoS. The generalized form of the
Chaplygin gas presents the mathematical formulation in which surface
pressure depends on the radius of the shell.

Therefore, we consider general form of Chaplygin EoS
($p=p(\sigma,b)$) to study the stable behavior of the respective
geometry, i.e., $p=\frac{1}{b^n}\frac{\eta}{\sigma}$ with real
constants $\eta<0$ and $n$ \cite{18}. It is observed that the
Chaplygin gas model is recovered for $n=0$ \cite{33}. The respective
solution of conservation equation for such a model can be written as
\begin{equation}\label{35}
\sigma^2=\frac{(n-4)\sigma_{0}^2b_{0}^{n+4}b^n+4\eta
b^4b_{0}^n-4\eta b^nb_{0}^4}{b^{n+4}b_{0}^{n}(n-4)}.
\end{equation}
The effective potential for this model turns out to be
\begin{eqnarray}\nonumber
\Omega(b)&=&-\frac{4 \pi ^2 b^{-n-2} b_0^{-n} \left(b_0^4 b^n
\left((n-4) \sigma_{0}^2 b_0^n-4 \eta\right)+4 b^4 \eta
b_0^n\right)}{n-4}+\frac{\zeta(b)}{2}\\\label{36}&-&\frac{(n-4)
b^{n+2} \xi(b)^2 b_0^n}{64 \pi ^2 \left(b_0^4 b^n \left((n-4)
\sigma_{0}^2 b_0^n-4 \eta\right)+4 b^4 \eta b_0^n\right)}.
\end{eqnarray}
It is observed that $\Omega(b_0)=0$ and $\Omega'(b_0)$ becomes
\begin{eqnarray}\nonumber
\Omega'(b_0)&=&-\frac{\eta b_0^{-n-3} \xi(b_0)^2}{16 \pi ^2
\sigma_{0}^4}+16 \pi ^2 \eta b_0^{1-n}-\frac{\xi(b_0) \xi'(b_0)}{32
\pi ^2 \sigma_{0}^2 b_0^2}-\frac{\xi(b_0)^2}{32 \pi ^2 \sigma_{0}^2
b_0^3}\\\nonumber&+&8 \pi ^2 \sigma_{0}^2 b_0+\frac{\zeta'(b_0)}{2}.
\end{eqnarray}
For $\Omega'(b_0)=0$, we have
\begin{equation}\label{37}
\eta=-\frac{\sigma_{0}^2 b_0^n \left(256 \pi ^4 \sigma_{0}^4
b_0^4+16 \pi ^2 \sigma_{0}^2 b_0^3 \zeta'(b_0)-b_0 \xi(b_0)
\xi'(b_0)-\xi(b_0)^2\right)}{2 \left(256 \pi ^4 \sigma_{0}^4
b_0^4-\xi(b_0)^2\right)}.
\end{equation}
Consequently,  $\Omega''(b_0)$ has the following form
\begin{eqnarray}\nonumber
\Omega''(b_0)&=&-\frac{\eta^2 b_0^{-2 (n+2)} \xi(b_0)^2}{2 \pi ^2
\sigma_{0}^6}-\frac{\eta b_0^{n-2 (n+2)+1} \xi(b_0) \xi'(b_0)}{4 \pi
^2 \sigma_{0}^4}+\frac{\eta n b_0^{n-2 (n+2)} \xi(b_0)^2}{16 \pi ^2
\sigma_{0}^4}\\\nonumber&-&\frac{7 \eta b_0^{n-2 (n+2)}
\xi(b_0)^2}{16 \pi ^2 \sigma_{0}^4}-16 \pi ^2 \eta b_0^{n-2
(n+2)+4}-16 \pi ^2 \eta n b_0^{n-2
(n+2)+4}\\\nonumber&-&\frac{b_0^{2 n-2 (n+2)+2} \xi(b_0)
\xi''(b_0)}{32 \pi ^2 \sigma_{0}^2}-\frac{b_0^{2 n-2 (n+2)+1}
\xi(b_0) \xi'(b_0)}{8 \pi ^2 \sigma_{0}^2}\\\nonumber&-&\frac{b_0^{2
n-2 (n+2)+2} \xi'(b_0)^2}{32 \pi ^2 \sigma_{0}^2}-\frac{b_0^{2 n-2
(n+2)} \xi(b_0)^2}{32 \pi ^2 \sigma_{0}^2}-24 \pi ^2 \sigma_{0}^2
b_0^{2 n-2 (n+2)+4}\\\label{38}&+&\frac{1}{2} b_0^{2 n-2 (n+2)+4}
\zeta''(b_0).
\end{eqnarray}

\begin{figure}\centering
\epsfig{file=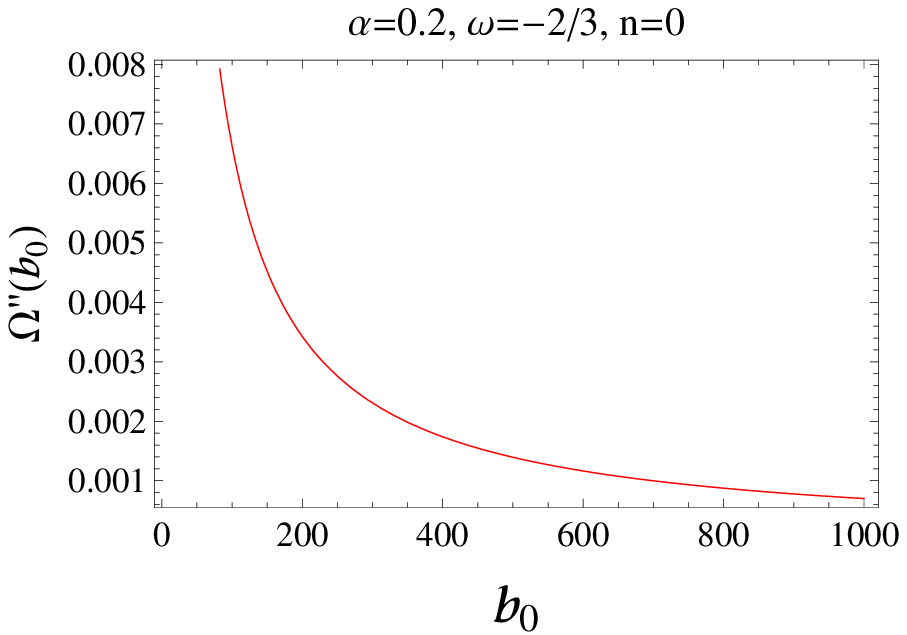,width=.5\linewidth}\epsfig{file=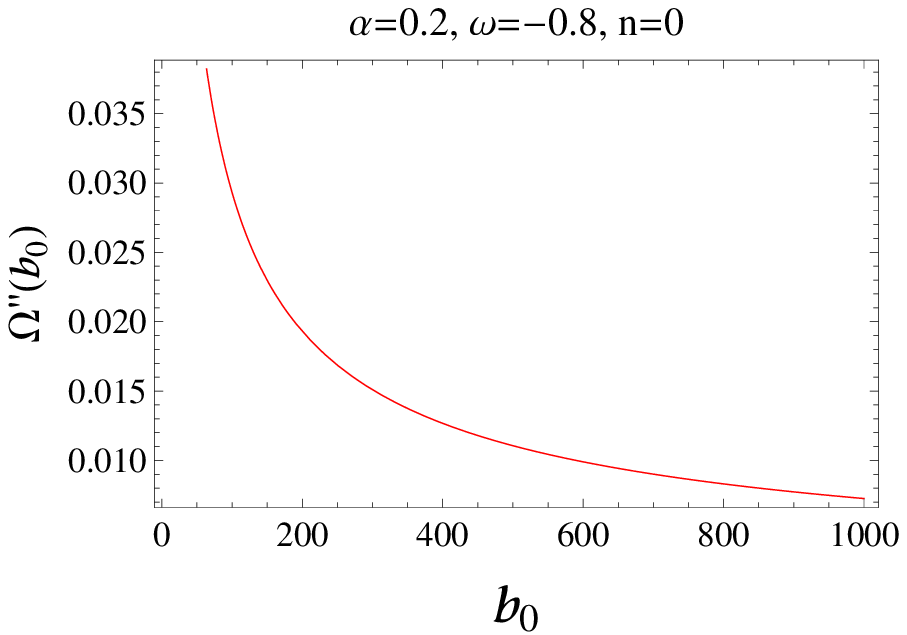,width=.5\linewidth}
\caption{Stable behavior of thin-shell gravastars with Chaplygin gas
model ($n=0$) for $\beta=0.5=m=Q$ with different values of
$\omega$.} \label{f3}
\end{figure}

\begin{figure}\centering
\epsfig{file=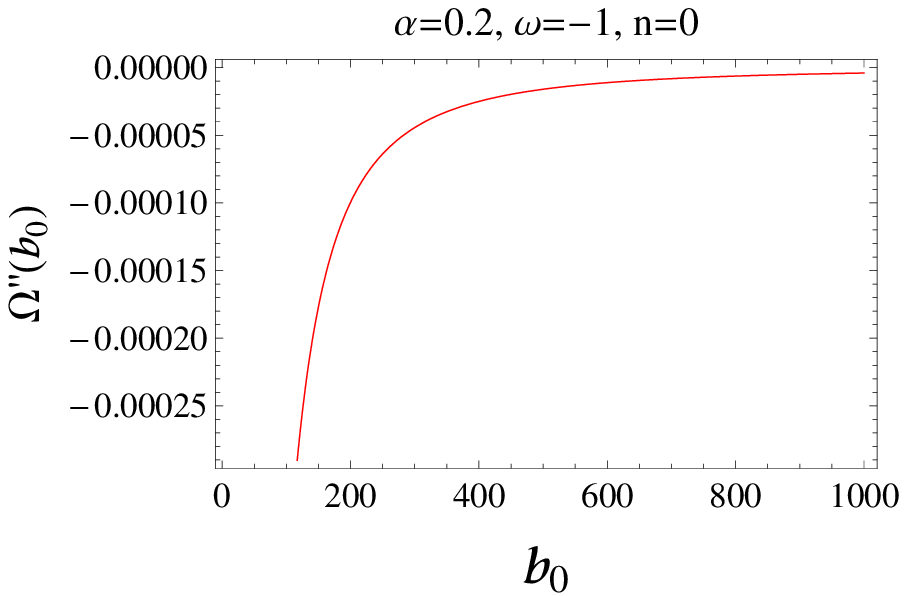,width=.5\linewidth}\epsfig{file=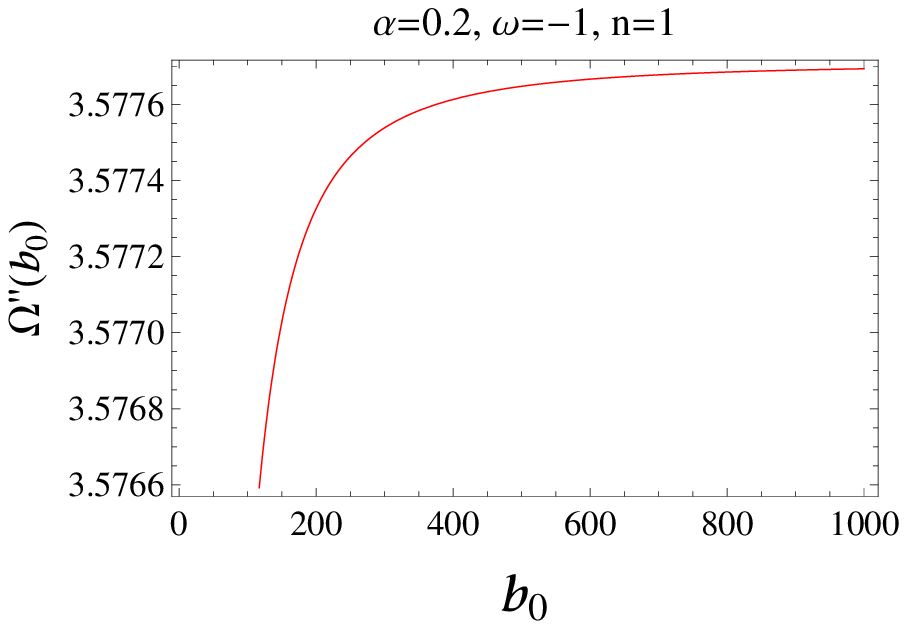,width=.5\linewidth}
\caption{Stability of thin-shell gravastars with generalized
Chaplygin gas EoS with different values of $n$. For $\omega=-1$, the
left plot shows unstable behavior for $n=0$ and right plot expresses
the stable structure for $n=1$.}\label{f4}
\epsfig{file=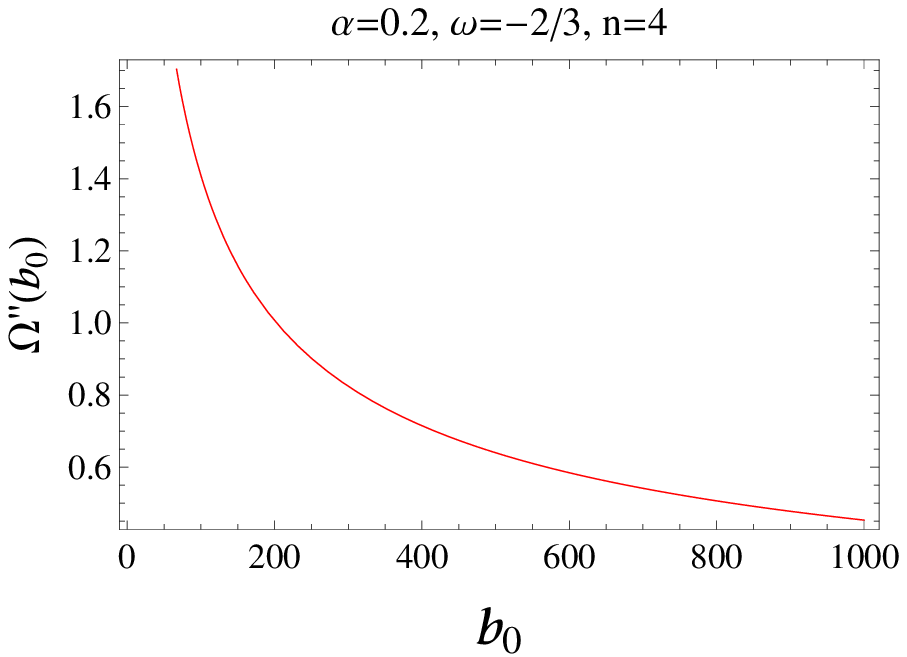,width=.5\linewidth}\epsfig{file=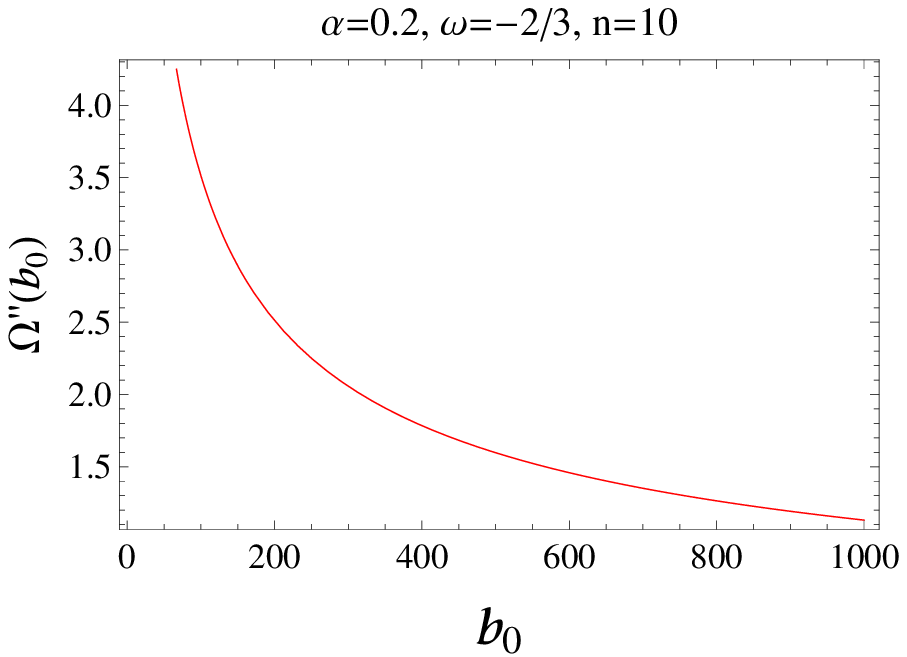,width=.5\linewidth}
\caption{Stable behavior of thin-shell gravastars for different
values of $n$. The stability of developed structure is enhanced for
large values of $n$.}\label{f5}
\end{figure}

Now, we observe the effects of the generalized Chaplygin gas EoS on
the stability of developed geometry. In this regard, we observe the
graphical behavior of $\Omega''(b_0)$ for this model. It is found
that thin-shell expresses stable behavior for every choice of the
physical parameters except $\omega=-1$ when $n=0$ (Figure
\ref{f3}). This shows that thin-shell becomes stable for the
choice of Chaplygin gas model ($n=0$) and represents unstable
behavior only for $\omega=-1$ (left plot of Figure \ref{f4}). It
is also analyzed that the general case of Chaplygen EoS ($n\neq0$)
shows stable behavior for every choice of $\omega$ with $n=1$ (right
plot of Figure \ref{f4}). We see that stable behavior
($\Omega''(b_0)>0$) increases for higher values of $n$ as shown in
Figure \ref{f5}.

\begin{figure}\centering
\epsfig{file=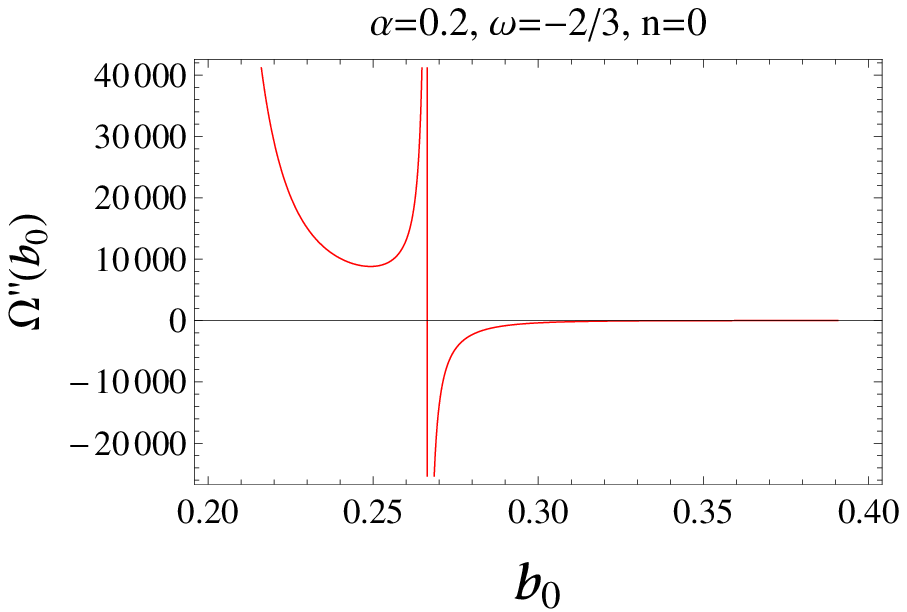,width=.5\linewidth}\epsfig{file=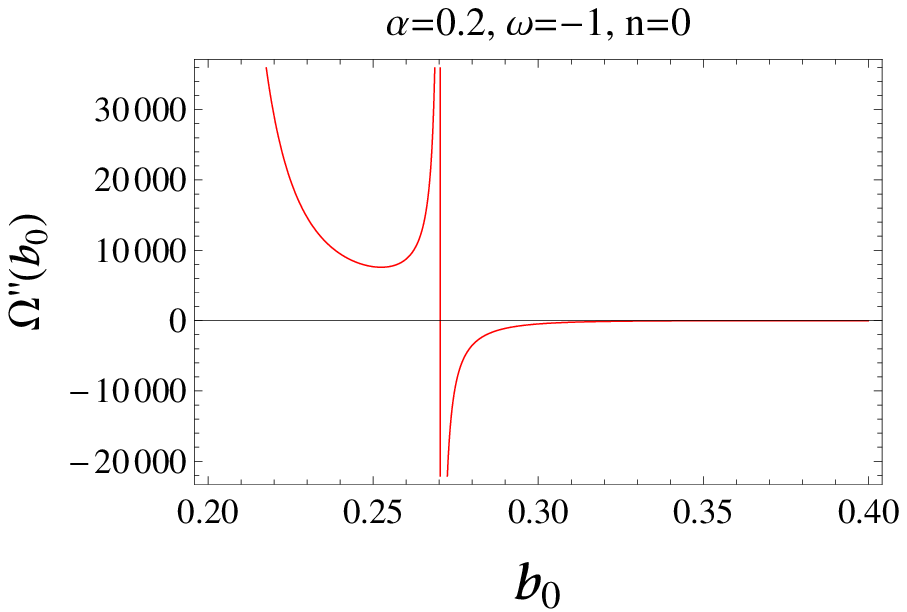,width=.5\linewidth}
\caption{Stable and unstable behavior of thin-shell gravastars with
phantomlike EoS ($n=0$) for different values of $\omega$. It shows
stable behavior initially then expresses unstable configuration for
every choice of $\omega$.}\label{f6}
\epsfig{file=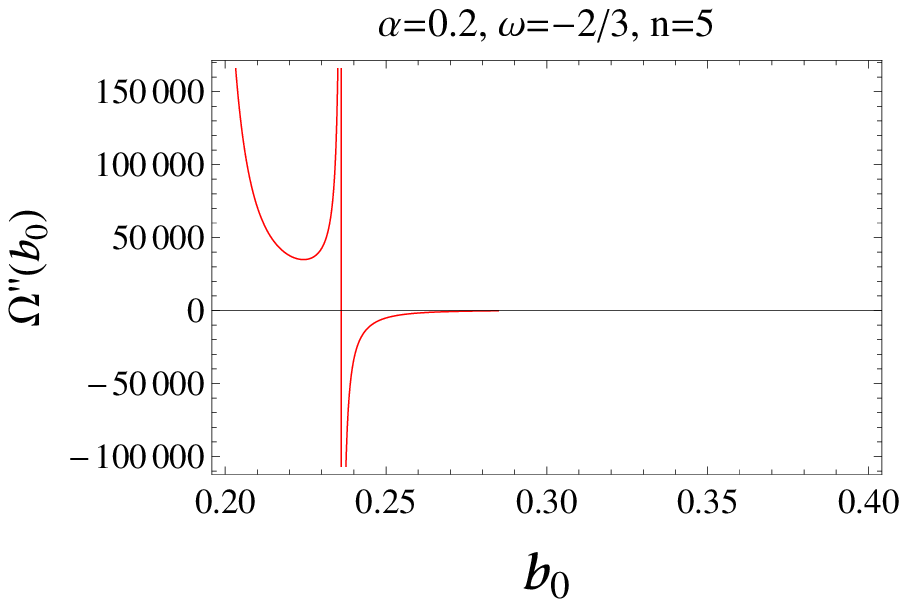,width=.5\linewidth}\epsfig{file=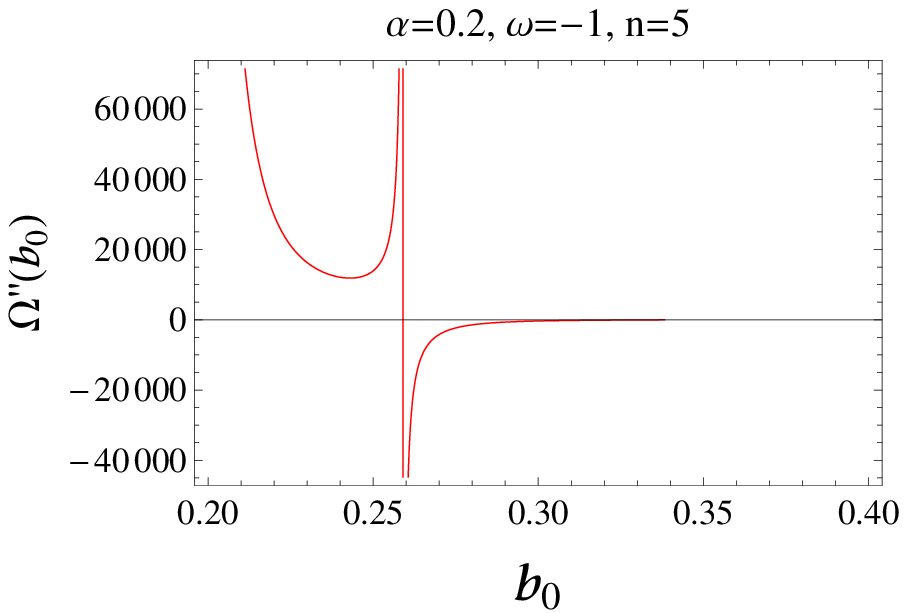,width=.5\linewidth}
\caption{Stable and unstable behavior of thin-shell gravastars for
generalized phantomlike EoS with different values of $\omega$.}\label{f7}
\end{figure}

Finally, we study the effects of generalized phantomlike variable
EoS on the stability of thin-shell \cite{18} whose EoS is
$p=\frac{\Theta\sigma}{b^n}$ with real constants $\Theta$ and $n$.
The phantomlike EoS is recovered if $n=0$ \cite{34}. By using this
expression in Eq.(17), we have
\begin{equation}\label{39}
\sigma=b_{0}^2b^{-2}\sigma_{0}e^{\frac{
\Theta\left(b_0^{-n}-b^{-n}\right)}{n}},
\end{equation}
and it follows that
\begin{equation}\label{40}
\Omega(b)=-\frac{b^2 \xi(b)^2 e^{\frac{2 \Theta
\left(b_0^{-n}-b^{-n}\right)}{n}}}{64 \pi ^2 \sigma_{0}
b_0^4}-\frac{4 \pi ^2 \sigma_{0} b_0^4 e^{\frac{2 \Theta
\left(b^{-n}-b_0^{-n}\right)}{n}}}{b^2}+\frac{\zeta(b)}{2}.
\end{equation}
It is noted that $\Omega(b_0)=0$ and by considering
$\Omega'(b_0)=0$, we obtain
\begin{equation}\label{41a}
\Theta=-\frac{b_0^n \left(256 \pi ^4 \sigma_{0}^2 b_0^4+16 \pi ^2
\sigma_{0} b_0^3 \zeta'(b_0)-b_0 \xi(b_0)
\xi'(b_0)-\xi(b_0)^2\right)}{256 \pi ^4 \sigma_{0}^2
b_0^4-\xi(b_0)^2},
\end{equation}
and hence
\begin{eqnarray}\nonumber
\Omega''(b_0)&=& -\frac{\Theta^2 b_0^{-2 n-4} \xi(b_0)^2}{16 \pi ^2
\sigma_{0}}-16 \pi ^2 \Theta^2 \sigma_{0} b_0^{-2 n}-\frac{\Theta
b_0^{-n-3} \xi(b_0) \xi'(b_0)}{8 \pi ^2
\sigma_{0}}\\\nonumber&-&\frac{\Theta b_0^{-n-4} \xi(b_0)^2}{16 \pi
^2 \sigma_{0}}-\frac{\Theta (1-n) b_0^{-n-4} \xi(b_0)^2}{32 \pi ^2
\sigma_{0}}-16 \pi ^2 \Theta \sigma_{0} b_0^{-n}\\\nonumber&+&8 \pi
^2 \Theta (-n-3) \sigma_{0} b_0^{-n}-\frac{\xi(b_0) \xi''(b_0)}{32
\pi ^2 \sigma_{0}b_0^2}-\frac{\xi(b_0) \xi'(b_0)}{8 \pi ^2
\sigma_{0} b_0^3}-\frac{\xi'(b_0)^2}{32 \pi ^2 \sigma_{0}
b_0^2}\\\label{41}&-&\frac{\xi(b_0)^2}{32 \pi ^2 \sigma_{0}
b_0^4}-24 \pi ^2 \sigma_{0}+\frac{\zeta''(b_0)}{2}.
\end{eqnarray}
For the general form of phantomlike EoS, we see that thin-shell
shows initially stable behavior then expresses unstable
configuration for every choice of physical parameters (Figures
\ref{f6} and \ref{f7}). We conclude that the constructed
geometry is neither stable nor unstable completely for the choice of
both phantomlike and general form of phantomlike EoS.

\section{Final Remarks}

This paper investigates the construction of thin-shell gravastars
from the matching of two different spacetimes, i.e., DS as a lower
spacetime and charged Kiselev BH as an upper manifold. These
geometries are connected through the well-known cut and paste
method. We match these manifolds at $r=b$ with $b>r_h$ to avoid the
presence of event horizon ($r_h$) and singularity in the developed
structure. The presence of matter thin layer at the joining surface
produces discontinuity in the extrinsic curvature. It is found that
the null energy condition is verified for the developed structure
(Figure \ref{f1}). We have studied stable characteristics of
thin-shell gravastars with barotropic type fluid distribution and
two variable EoS, i.e., generalized Chaplygin gas and phantomlike
EoS.

For barotropic model, we have obtained stable solution for the
choice of $\omega=-1$ and unstable solution for any other choice of
$\omega$ (Figure \ref{f2}). It is interesting to mention here that
this model mostly indicates unstable behavior for thin-shell WHs in
several spacetimes \cite{18,19,31,31a}. These results express that
the stable solution can be obtained through barotropic model for
some suitable choice of physical parameters. The stable structure is
obtained for Chaplygin gas model ($n=0$) for every choice of
$\omega$ other than extraordinary quintessence parameter
$\omega=-1$. For generalized Chaplygin gas EoS, we have obtained
stable solution for every values of physical parameter and found
more stable structure for higher values of $n$ (Figures
\ref{f3}, \ref{f4} and \ref{f5}. Finally, for generalized phantomlike EoS, thin-shell
shows initially stable behavior and then expresses unstable
configuration for every choice of the physical parameters (Figures
\ref{f6} and \ref{f7}).

We conclude that charged Kiselev thin-shell gravastars are more
stable for the choice of generalized Chaplygin gas model. It is
worthwhile to mention here that this model is more stable with
considered EoS than thin-shell WHs in the background of various BHs
\cite{18,19,31,31a}. This shows completely stable structure of
thin-shell gravastar with extraordinary quintessence parameter for
both barotropic and generalized Chaplygin gas model.

\section*{Appendix A}

We also explore some physical features of the developed structure,
i.e., proper length, entropy and energy contents within the shell's
region. Since the constructed geometry is the matching of two
different spacetimes, so the stiff perfect fluid moves along these
spacetimes through the shell region. The lower and upper boundaries
of the shell are $r=b$ and $r=b+\epsilon$, respectively. The proper
thickness of the shell is denoted by $\varepsilon$ which is a very
small positive real number ($0<\varepsilon\ll1$). The proper
thickness of such a region that connects lower and upper spacetimes
can be obtained as \cite{13}
\begin{equation}\label{19}
l=\int_{b}^{b+\epsilon}\sqrt{\Psi^{-1}(r)}dr=\int_{b}^{b+\epsilon}
\frac{dr}{\sqrt{1-\frac{2m}{r}-\frac{\alpha}
{r^{3\omega+1}}+\frac{Q^2}{r^2}}}.
\end{equation}
This integral cannot be solved analytically due to the complicated
expression of $\Psi(r)$. Therefore, we solve it by assuming
$\sqrt{\Psi^{-1}_{+}(r)}=\frac{dj(r)}{dr}$ as
\begin{equation}\label{20}
l=\int_{b}^{b+\epsilon}\frac{dj(r)}{dr}dr=j(b+\epsilon)-j(b)\approx
\epsilon \frac{dj(r)}{dr}|_{r=b}=\epsilon\sqrt{\Psi^{-1}_{+}(b)},
\end{equation}
where $\epsilon\ll1$ so that its square and higher powers can be
neglected. The corresponding expression for proper length becomes
\begin{equation}\label{21}
l=\epsilon\left[1-\frac{2m}{b}-\frac{\alpha}
{b^{3\omega+1}}+\frac{Q^2}{b^2}\right]^{-\frac{1}{2}}.
\end{equation}
It is noted that the proper length of the shell clearly depends on
the charge as well as the mass of the BH. Equation (\ref{21}) shows
that the proper length and thickness of the shell are proportional.
It is found that the length of the shell decreases by an increasing
charge of the geometry and increases by increasing the mass of the
BH.

\begin{figure}\centering
\epsfig{file=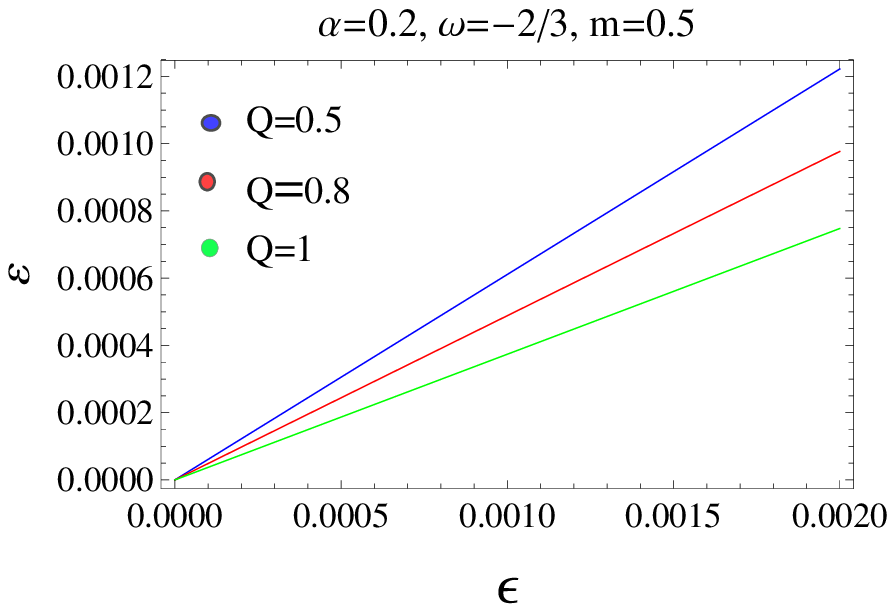,width=.5\linewidth}\epsfig{file=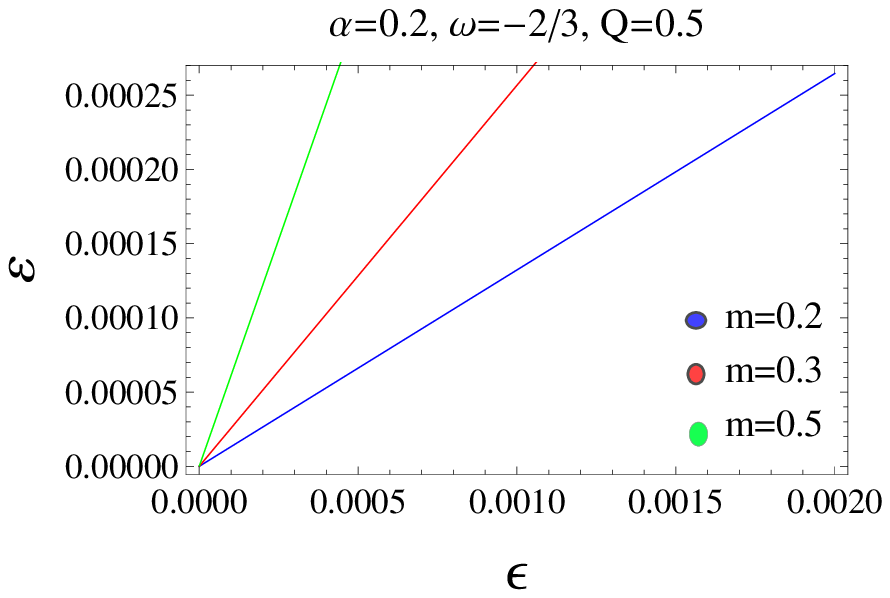,width=.5\linewidth}
\caption{Behavior of the energy within the shell verses thickness of
the shell with $\beta=0.5$ and $b_0=1$.}\label{f9}
\end{figure}

Entropy is related to the measure of disorderness or disturbance in
a geometrical structure. We study the entropy of thin-shell
gravastars that explains the disorderness in the geometry of
gravastar. According to the theory of Mazur and Mottola, charged
gravastar has zero entropy density for the interior region. Using
the concept of Mazur and Mottola, we evaluate the entropy of
thin-shell gravastar through the expression \cite{13}
\begin{equation}\label{22}
S=\int_{b}^{b+\epsilon}4\pi r^2 h(r) \sqrt{\Psi^{-1}(r)}dr.
\end{equation}
The entropy density for local temperature can be expressed as
\begin{equation}\label{23}
h(r)=\frac{\vartheta K_B}{\hbar}\sqrt{\frac{p(r)}{2\pi}},
\end{equation}
where $\vartheta$ is a dimensionless parameter. Here, we take Planck
units $(K_B = 1 = \hbar)$ so that the shell's entropy becomes
\cite{13}
\begin{figure}\centering
\epsfig{file=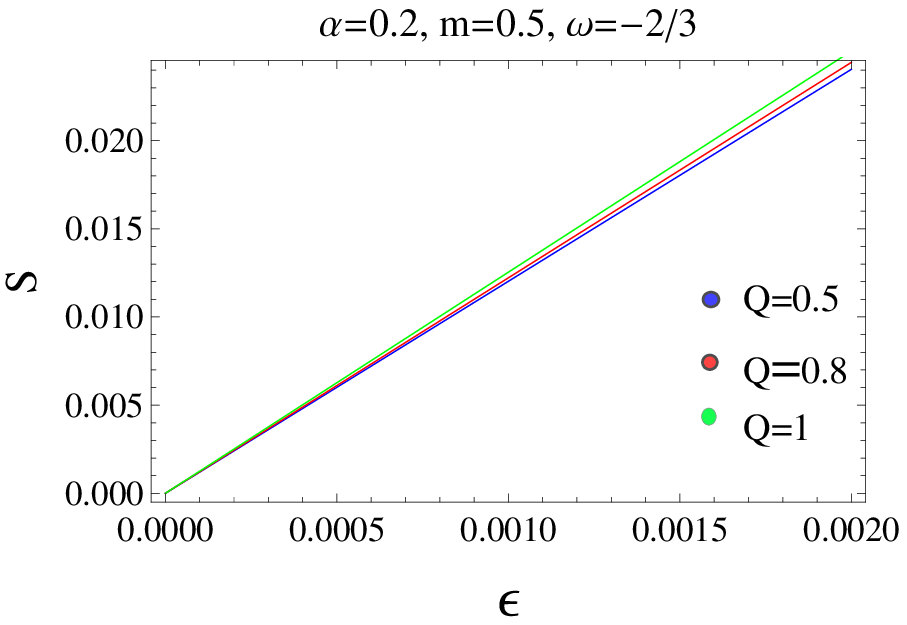,width=.5\linewidth}\epsfig{file=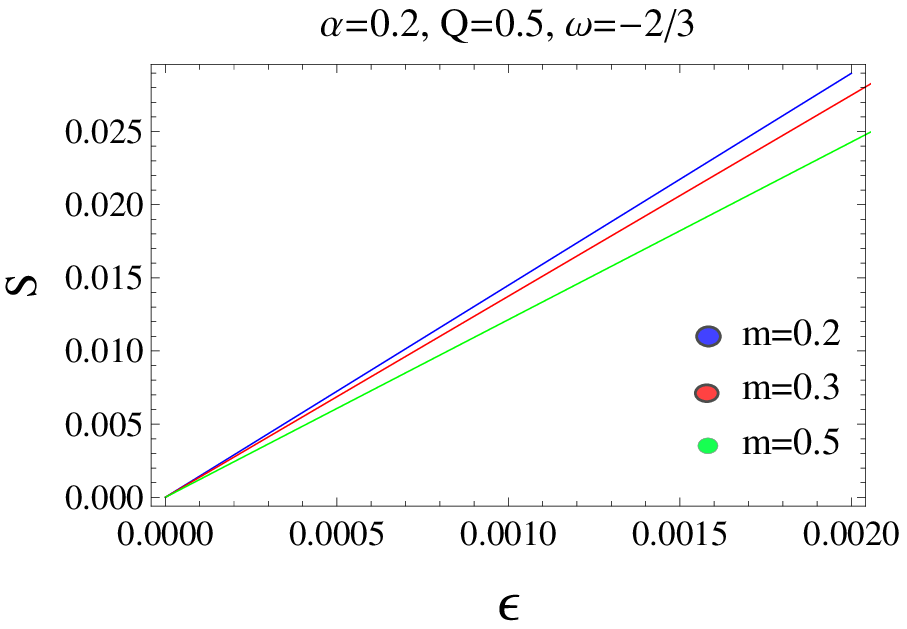,width=.5\linewidth}
\caption{Behavior of entropy versus thickness of the shell with
$\beta=0.5$ and $b_0=1$.}\label{f8}
\end{figure}
\begin{equation}\label{24}
S=\epsilon \vartheta b^2\sqrt{8\pi p(b)\Psi^{-1}(b)}.
\end{equation}
It is shown that entropy of the shell's region is also proportional
to the shell's thickness. We use this equation to examine the
contribution of charge and mass of BH on the entropy of shell
graphically. Figure \ref{f8} shows the linear relation between
entropy and thickness for different values of the physical
parameter. It is found that the entropy of shell region increases by
increasing $Q$ and decreases for large values of $m$. The interior
region of gravastars obeys the EoS $p=-\sigma$ which represents
negative energy zone with non-attractive force. The energy
distribution in the shell's region can be determined as \cite{13}
\begin{equation}\label{25}
\varepsilon=\int_{b}^{b+\epsilon}4\pi r^2
\sigma(r)dr\approx4\epsilon\pi b^2 \sigma(b).
\end{equation}
The energy contents depend on the thickness of the shell, mass and
charge of the geometry. We see that energy within the shell
decreases for large values of charge and increases for large values
of mass as shown in Figure \ref{f9}.

It is concluded that these features are proportional to the
thickness of the shell and are greatly affected by the charge and
mass of the BH which is consistent with the literature
\cite{13}-\cite{16}.

\section*{Acknowledgement}

One of us (FJ) would like to thank the Higher Education Commission,
Islamabad, for its financial support through
\emph{6748/Punjab/NRPU/RD/HEC/2016}.


\begin{thebibliography}{43}

\bibitem{1} Mazur, P. and Mottola, E.: Report No. LA-UR-01- 5067 (arXiv:gr-qc/0109035).

\bibitem{2} Mazur, P. and Mottola, E.: Proc. Nat. Acad. Sci.
\textbf{101}(2004)9545.

\bibitem{2a} Visser, M., Kar, S. and Dadhich, N.: Phys. Rev. Lett.
\textbf{90}(2003)201102.

\bibitem{3} Visser, M. and Wiltshire, D.L.: Class. Quantum Grav.
\textbf{21}(2004)1135.

\bibitem{4} Carter, B.M.N.: Class. Quantum Grav.
\textbf{22}(2005)4551.

\bibitem{5} Horvat, D., Sasa Ilijic, S. and Marunovic, A.:
Class. Quantum Grav. \textbf{26}(2009)025003.

\bibitem{6} Rahaman, F., Usmani, A.A., Ray, S. and Islam, S.: Phys. Lett.
B \textbf{707}(2012)319; ibid. \textbf{717}(2012)1.

\bibitem{7} Banerjee, A., Rahaman, F., Islam, S. and Govender, M.: Eur.
Phys. J. C \textbf{76}(2016)34.

\bibitem{8} Rocha, P., Chan, R., da Silva, M.F.A. and Wang, A.: J. Cosmol. Astropart. Phys.
\textbf{2008}(2008)10; ibid. \textbf{2009}(2009)10; ibid.
\textbf{2011}(2011)13.

\bibitem{9} Horvat, D., Ilijic, S. and Marunovic, A.: Class. Quantum Grav.
\textbf{28}(2011)195008.

\bibitem{10} Lobo, F.S.N. and Garattini, R.: J. High Energy Phys.
\textbf{1312}(2013)065.

\bibitem{11} \"{O}vg\"{u}n, A., Banerjee, A. and Jusufi, K.: Eur. Phys. J. C
\textbf{77}(2017)566.

\bibitem{12} Sharif, M. and Javed, F.:
doi.org/10.1016/j.aop.2020.168124.

\bibitem{13} Ghosh, S., Rahaman, F., Guha, B.K. and Ray, S.: Phys. Lett. B
\textbf{767}(2017)380.

\bibitem{14} Shamir, M.F. and Ahmad, M.: Phys. Rev. D
\textbf{97}(2018)104031.

\bibitem{16} Yousaf, Z. et al.: Phys. Rev. D \textbf{100}(2019)024062.

\bibitem{17} Sharif, M. and Waseem, A.: Astrophys. Space Sci.
\textbf{364}(2019)189.

\bibitem{18} Varela, V.: Phys. Rev. D \textbf{92}(2015)044002.

\bibitem{19} Sharif, M. and Javed, F.: Gen. Relativ. Gravit.
\textbf{48}(2016)158; Astrophys. Space Sci. \textbf{364}(2019)179.

\bibitem{20} N$\acute{u}\tilde{n}$ez, D., Quevedo, H. and Salgado, M.: Phys. Rev. D
\textbf{58}(1998)083506.

\bibitem{22} Mazharimousavi, S.H., Halilsoy, M. and Amirabi, Z.:
Phys. Rev. D \textbf{81}(2010)104002.

\bibitem{23} Rahaman, F., Ray, S., Jafry, A.K. and Chakraborty, K.: Phys. Rev. D
\textbf{82}(2010)104055.

\bibitem{24} Dias, G.A.S. and Lemos, J.P.S.: Phys. Rev. D
\textbf{82}(2010)084023.

\bibitem{25} Sharif, M. and Abbas, G.: Gen. Relativ.
Gravit. \textbf{43}(2011)1179.

\bibitem{26} Sharif, M. and Azam, M.: Eur. Phys. J. C
\textbf{73}(2013)2407.

\bibitem{27} Rahaman, F., Banerjee, A. and Radinschi, I.: Int. J. Theor. Phys.
\textbf{52}(2013)2943.

\bibitem{28} Sharif, M. and Iftikhar, S.: Astrophys. Space Sci.
\textbf{356}(2015)89.

\bibitem{29} Forghani, S.D., Habib Mazharimousavi, S. and
Halilsoy, M.: Eur. Phys. J. C \textbf{78}(2018)469.

\bibitem{30} Sharif, M. and Javed, F.: Int. J. Mod. Phys. D
\textbf{28}(2019)1950046; Ann. Phys. \textbf{407}(2019)198; Mod.
Phys. Lett. A \textbf{35}(2019)1950350; Chin. J. Phys.
\textbf{61}(2019)262.

\bibitem{31} Kuhfittig, P.K.F.: Turk. J. Phys. \textbf{43}(2019)213.

\bibitem{31a} Sharif, M. and Javed, F.: doi.org/10.1142/S0217751X20400151; Int. J. Mod. Phys. D
\textbf{29}(2020)2050007; doi.org/10.1016/j.aop.2020.168146; Int. J.
Mod. Phys. A \textbf{35}(2020)2050030;
doi.org/10.1142/S0217732320503095.

\bibitem{31b} Perlmutter, S. et al.: Astrophys. J.
\textbf{517}(1999)565.

\bibitem{31c} Hellerman, S., Kaloper, N. and Susskind, L.: J. High Energy Phys.
\textbf{6}(2001)3.

\bibitem{32} Kiselev, V.V.: Class. Quantum
Grav. \textbf{20}(2003)1187.

\bibitem{32a} Riess, A. et al.: Astron. J. \textbf{116}(1998)1009;
Perlmutter, S.J. et al.: Astroph. J. \textbf{517}(1999)565; Bahcall,
N.A. et al.: Science \textbf{284}(1999)1481.

\bibitem{32b} Sahni, V. and Starobinsky, A.A.: Int. J. Mod. Phys. A \textbf{9}(2000)373;
Peebles, P.J. and Ratra, B.: Rev. Mod. Phys. \textbf{75}(2003)559;
Padmanabhan, T.: Phys. Rep. \textbf{380}(2003)235.

\bibitem{32c} Kamenshchik, A., Moschella, U. and Pasquier, V.: Phys. Lett. B \textbf{511}(2001)265;
Bili´c, N., Tupper, G.B. and Viollier, R.D.: Phys. Lett. B
\textbf{535}(2002)17; Bento, M.C., Bertolami, O. and Sen, A.A.:
Phys. Rev. D \textbf{66}(2002)043507.

\bibitem{33} Eiroa, E.F. and Simeone, C.: Phys. Rev. D
\textbf{76}(2007)024021.

\bibitem{34} Kuhfittig, P.K.F.: Acta Phys. Polon. B \textbf{41}(2010)2017.

\end{thebibliography}
\end{document}